\documentclass[12pt]{article}                    
\usepackage{graphicx,natbib,hyperref}
 \usepackage{mathptmx}      
\usepackage{amsmath,url,geometry}
\usepackage[utf8]{inputenc}
\def\email#1{\href{mailto:#1}{#1}}
\begin{document}
\title{Protention and retention in biological systems\footnote{This paper was made possible
only as part of an extended
collaboration with Francis Bailly (see references), a dear friend and
``ma\^itre \`a penser'', who contributed to the key ideas. Francis passed
away in february 2008: we continue here our inspiring discussions and
joint work.}}


\author{ Giuseppe Longo\footnote{Informatique, CNRS – Ecole Normale Sup\'erieure and CREA,Paris 
              \email{longo@di.ens.fr}
	      \url{http://www.di.ens.fr/users/longo}} \and Maël Montévil\footnote{Mathématiques, Ecole Normale Sup\'erieure and ED Frontières du
vivant, Paris V, Paris
\email{montevil@di.ens.fr}}}




\maketitle

\paragraph{Abstract}
This paper\footnote{Published in
\emph{Theory in Biosciences}, vol. 130, n.2, pp. 107–117, 2011. doi:
\href{http://dx.doi.org/10.1007/s12064-010-0116-6}{10.1007/s12064-010-0116-6}.
} proposes an abstract mathematical frame for describing some features
of cognitive and biological time. We focus here on the so called ``extended
present'' as a result of protentional and retentional activities (memory and
anticipation). Memory, as retention, is treated in some physical theories
(relaxation phenomena, which will inspire our approach), while protention (or
anticipation) seems outside the scope of
physics. We then suggest a
simple functional representation of biological protention. This allows us to
introduce the abstract notion of “biological inertia”.

\section*{Introduction}
\noindent The notions of “memory” and “anticipation” are analyzed here from a
temporal
perspective. By this, we propose a simple mathematical approach
to \emph{retention} and \emph{protention} that are apparently shared by all
organisms, albeit rudimentarily.  Moreover, in life phenomena, memory is
essential to learning and it is
oriented towards action, the grounding of protention. Our approach will allow
to address the issue of what
we call “biological inertia”, a form of ``continuation'' of ongoing
action, derived from the notions above. The frame is purely mathematical and
abstract: only practitioners will be able to give values to our coefficients and
develop, possibly, concrete applications of the approach, from cell biology to
human cognition. Our aim is to give a precise and relevant meaning to notions
that
are usually treated in a rather informal fashion and unrelated between one
another, such as those of time of representation, time of retention and time of
protention.

A long phenomenological tradition
introduces an important distinction between memory and retention, on the one
hand, and anticipation and protention on the other. In short, the common meaning
of “memory” seems to essentially refer to a “conscious reconstruction” of
something that was experienced (very well put by Edelman as a “brain which sets
itself back into a previously experienced state”). Anticipation would be its
temporal opposite --- the awareness of an expectation, of a possible future
situation. Memory and anticipation do not, a priori, have a biological
\emph{characteristic time}, a notion which is essential to our analysis. 
In our approach, instead, possibly pre-conscious retention is to be
seen as an extension of the present; it is the present which is “retained”,
during a brief interval of time (related to what will be called its
characteristic time) for the objective of the action (and of perception), it is a form of
extension of the immediate past into the present. For example, when listening to
a word or a phrase, we retain the part which has already
occurred for a certain (characteristic) duration of time. The mental
duration of a phrase, particularly of a musical “phrase”, is needed for grasping
meaning or a melody (see for example \cite{Perfetti1976, Nicolas}): it is the
present which leaves a trace the time necessary
for action or, possibly, for subsequent awareness. 
But protention (as preconscious anticipation) is essential to appreciate
a melody or understand a phrase. When reading, the analysis of saccadic eye
movements demonstrates that we first look
at least at half of the word following the one we are reading, see \cite{wildman1978semantic}. This protentional
behaviour participates in the reconstruction of meaning: we appear to make sure
of the meaning of the word we are reading by making a partial guess upon the
following word.

Technically, protention will be given by a temporal mirror
image, as it extends retention forwards into time. Protention is, above all,
the \emph{tropism} inherent to action performed by any life form. This point is
at the center of our approach: we call retention and protention these particular
aspects of memory and of anticipation that are specific to all life forms -- a
sort of present which is extended in both directions. Thus we do not limit our
analysis to the phenomenological use of these words, inasmuch it limits their
meaning
to situations that can be examined through conscious activities. We believe that
this extension to pre-conscious activities
remains compatible with (and helps to understand) its classical usage,
particularly such as described by \cite{vangelder1999} and \cite{varela1999} who
develop the concepts of intentionality, retention and of protention, introduced
and discussed in length by Husserl in his analysis of human consciousness.

In this paper, it is then a question of trying to propose an
elementary modelization of these inevitably fuzzy notions, one which is as
rudimentary as possible, but one that can nevertheless support discussions  regarding their precise conceptualization and their increasingly
thorough mathematization. The introduction of the notions of ``biological
inertia'' and  ``global protention'' are, typically, a consequence of the generative power of mathematics.

To our aim, we will define some basic principles and more specific notions, after
some methodological preliminaries.

\subsection*{Methodological remarks}
\noindent This paper belongs to series of texts \cite{bailly2010l,bailly2008,bailly2009,bailly2010} whose
attempted aim is not to reconstruct the physico-mathematical complexity of
some aspects of biology, but to propose firstly and above all a proper biological
perspective. We believe that the \emph{theoretical differentiation} between
theories of inert and of living phenomena requires, among other things, a change
in the relevant \emph{parameters} and \emph{observables}. As long as the actions
of living organisms, including their cognitive performances which occur the
moment that life appears (in this sense, we speak of protention and of
retention in the amoeba or the paramecium), are analyzed within physical
space-time, the physico-mathematical takes precedent over the specificity of the
biological. For example, the formidable mathematics of morphogenesis, from
phyllotaxy to the analysis of the fractal structures of organs, organize the
results of friction in the growth of living organisms according to physical
geodesics. This friction is nevertheless shaped within physical space-time
(fractality optimizes the occupation of physical space, the exchange of energy
by a surface within a volume~\ldots). In all of these cases, the spatio-temporal
and energetic parameters and observables enable a very interesting and often
technically very difficult analysis. This is an approach of the \emph{physical
complexity} of living phenomena and of its material structures. We could also
say the same of analyses of networks of cells, of which the most complex are
neural networks. Informational interaction, often a gradient of energy, enables
to develop a theory, now very rich from the mathematical standpoint, of
these formal networks of which increasingly important applications are being
considered for the construction of machines that are somewhat intelligent (at
last).

In this paper, our mathematics will not go beyond a few equations
which could be presented to high school students. What matters in our view is
approaching biological time according to its own specificity, by starting with
some invariants which appear to be \emph{exclusively} specific to living
phenomena, as we did in \cite{bailly2010l}, or with properties that are not treated
by current physical theories, as protention here. In \cite{bailly2010}, we
proposed a two dimensional representation of biological time as a mathematical
frame to accomodate the autonomous (internal) biological rhythms (cardiac,
respiratory, metabolic rhythms~\dots). In the perspective of this paper, one
may understand the expectation or anticipation
of a rhythm to iterate, as a minimal form of protention: once rhythms are
installed, the organism is ``tuned'' to (and ``expects'') their iteration. 

Before developing a further geometrization of biological-time, we will face yet another
taboo of physicalism in
biology: the inverted causality specific to protention. We will not present a
physical theory of teleonomy, but will use as data the evidence of protentional
behaviours that may be observed in any life form. When the paramecium, encircled
by a ring of salt, tries after many attempts to break through the obstacle,
risking its own life and possibly even succeeding \cite{Misslin}, we can take note
of the retention/memory-learning of what we see and of the ensuing teleonomic
gesture (a protention) and develop an adequate theory (see \cite{amoeba}).
Likewise, when we hear that the brain, prior to a saccadic eye movement, in an
obvious anticipation, prepares the corresponding primary cortex which is apt to
receive the new signal (see \cite{Berthoz02}), there is certainly an underlying
physico-chemical mechanism which will one day enable to grasp the phenomenon by
means of physical causality, a causality which may need to be
invented. For the moment, let’s consider these phenomena as a
form of protention to be analyzed (correlated, formalized~\ldots) by a theory
specific to living phenomena, even if it has no correspondence or meaning within
current physical theories. Then, the unification with the
physico-chemical theories may be better considered, in order to evidentiate the physico-chemical components
which underlie these phenomena. As a matter of fact, unification will be possible only when we
will have \emph{two} theories to compare to one another, both theories being as
mathematized as possible. We are talking about \emph{unification} and not
\emph{reduction}, since physicists aim to unify the relativistic and quantum
fields and not perform a reduction of the one \emph{theory} to the other: string
theory and non-commutative geometry aims the construction of a new unified frame
which presents a new perspective for both theories.
The mathematics to be found in the following pages will give us the advantage of
formalization: it forces to specify concepts and to stabilize them as much as
possible (this is what mathematics is first about). Maybe that which follows is
false, but it should then be possible to say so in relation to a precise
formulation.

\section{Characteristic time and correlation lengths}
\noindent  The notion of “characteristic time”, which we inherit here from
physics, appears
to be very important in biology as well: it concerns the unity of the living
individual because, for example, fluxes and their transport entail lengths and,
therefore, relevant transport times. We will also speak of characteristic times
for retention and protention.

For example, according to the size of the organism, there appears to be two sorts of transport processes. For
large organisms, it would be of a “propagational” type ($v_p$ velocity, along
networks and ``channels'') with a
typical correlation length of $L_p = v_p\tau$, where $\tau$ represents the
characteristic time. For smaller organisms (cells, for example), it would rather
be of a “diffusional” type (diffusion coefficient $D$, due to
molecular diffusion processes) and the typical
correlation length would be $L_d = (D\tau)^\frac{1}{2}$.

We stress the difference regarding dependency in function of time: linear in one
case, as a power of $\frac{1}{2}$ in the other.

Two complementary remarks:\\
\begin{itemize}
\item The size of the organism also affects structures determining the mode of
transport, for example the respiratory function (oxygen transport): in the case
of small organisms (insects, for example) the transport is performed by
\emph{tracheas} (or even pores), multitudes of little cylinders where the air
diffuses in order to reach the cells. In the case of large organisms (fish,
mammals), transportation and exchanges are performed by means of \emph{gills} or
of \emph{lungs}, centralized anatomic structures which present the fractal
geometries we evoked above and which enable to conciliate difficultly compatible
constraints (efficiency, steric limitation, homogeneity), and then by various
sorts of vascular systems. Transportation, in
this last case, is also much more of a “propagational" type (even if diffusion
does play a role, namely in bronchioles).\\
\item These considerations essentially apply to various \emph{structural}
aspects responding to identical functions. The \emph{functional} aspect responds
for its part very generally to common scaling laws (the metabolism which
corresponds particularly to oxygen intake, the variegated rhythmicities, the
relaxation times~\ldots). It therefore appears that the modes of transport
associated to identical functions can be different and can correspond to
different anatomic structures (tracheas, gills, bronchial trees/lungs). This
is the well-known phenomenon of analogy of
structures in evolutionary biology.\\
\end{itemize}
Finally, account taken of these remarks, since the characteristic times $\tau$
mostly scale as $W_f^\frac{1}{4}$,  where $W_f$ is the mass of the intended
organism (see \cite{lindstedt,savage2004}), it
is necessary to expect the correlation lengths to scale differently according to
the mode of transport: respectively $L_p$  in  $W_f^\frac{1}{4}$  and  $L_d$  in
 $W_f^\frac{1}{8}$, following the definitions of  $L_p$  and $ L_d$.

In the sequel, our characteristic times will more precisely refer to
``relaxation times'', still in analogy to physics (see next footnote), yet in
properly biological frame, in relation to retention and protention.

\subsection{Critical states and correlation length}
\noindent The physics of criticality and self-organized systems has massively
entered the domain of biology since early ideas by \cite{Nicolis77},
 \cite{Bak88}, \cite{Kauffman93}~\dots\ 
We further extended this approach, in direct reference to far from equilibrium
systems in the sense of Prigogine, by considering living entities as being in an
``extended critical situation'', beyond the pointwise analysis of critical
transitions proper to physical theories, see \cite{bailly2008}.

It is interesting now to consider that
physical criticality is associated with a so-called critical slowdown (see for example \cite{suzuki1982critical}: the
relaxation time of a system tends to infinity when it goes near the critical
point. The qualitative meaning of these situations in biology is that the effect
of a stimuli would take a long time to stabilize (or, more generally, the organism would take a long time to
``react'' or ``adjust''), if one views life as close or in an (extended) 
critical state. In particular, criticality would lead to very slow cognitive
reactions if reaction needs a stabilization.

More generally, also in an information theoretic perspective, the
elaboration/reaction time is necessarely slow in an organism
with long correlations in space and slow characteristic time of the
individual components of the system.
However organisms and especially metazoans must often react quickly
and are able to do so. Consequently, biological organization provides a solution
to this paradox. This solution is to compensate this slowness by preparing the
organism to a forthcoming stimulus \emph{in advance}. We will try to
provide
a simple framework to tackle these properties, by an analysis of protention
and biological inertia. Of course, in this context,
perception itself is co-determined by this protentional activity.

\section{Retention and protention.}
\label{ret}

\subsection{Principles}

\noindent We therefore consider \emph{retention} $R$ by specifying it under the
form:
\begin{center}
$R_k(t_0,t)$ at an instant $ t$ of an anterior “event” $e$ of nature
$ k$ at time $t_0$,
\end{center}
For short and if needed, we will pose that  $e^k_0 = e^k(t_0)$  (where  $t_0
\leq t$).

Virtual protention, of an event of the same nature $e^k_1 = e^k_{t_1}$ at moment
$t$ of an ulterior instant $t_1$  $(t \leq t_1)$  will be noted $
V_{Pk}(t,t_1)$.  However, (actual) protention will be considered as a
function also of retention
$R_k$ because, and this is an essential principle of our approach, in \emph{the
absence of the retention of an event of nature $k$ there will be no possible
protention for an event of such nature}. We will therefore have
$P_k(R_k,t,t_1) = 0$,  for $ R_k = 0$. For the sake of simplicity, we
described this dependence of protention on retention as a linear dependence and
our (actual) protention, $P_k =  R_kV_{Pk}(t,t_1)$, will express
this\footnote{\label{maning} After reading a draft of this paper, L. Manning
gave us references to IRM data confirming the neurophysiological and
neuroimaging evidence for protention and the dependence of protention on
retention: \cite{szpunar2007,botzung2008}. Further, more specific experiments
would be required in order to quantify the coefficients we introduce here and
check/adjust the linearity of this dependence.}. Moreover, in conformity with
our previous analyses, we will pose that this  protention is a monotonous
increasing function of the retention in question, that is $\frac{\partial
P_k}{\partial R_k} \geq 0$.

\subsection{Specifications}
\noindent On the basis of the distinction made above, we have thus introduced
the notions
of retention and of virtual protention, as “immediate” and “passive” memory and
anticipation in order to express the fact that what we have are phenomena that
do not stem from the intentionality related to a conscious activity of a subject
(generally endowed with a more or less elaborate nervous system), but to simple
processes of biological reaction/stimuli/response, of which many primitive
organisms in relationship to their environment are the locus. To the end of
developing this point of view, we now introduce distinct concepts with effects
which we propose to represent by means of simple functions, mainly
\emph{relaxation functions} and their combinations\footnote{Relaxation functions
are among the simplest decreasing functions enabling to define a characteristic
time $\tau$ in physics, they often represent the basic model for the return to
the  equilibrium of a system that was initially brought out of equilibrium, with
the speed at which the system returns to the equilibrium $f_e$ of the system’s
$f$ function ($\frac{df}{dt}$) remaining proportional to this interval 
$\frac{df}{dt} = - \frac{|f-f_e|}{\tau}$.}.

More specifically, we will first define the retention function:\\
\begin{align}
R(t_0,t) = a_R\exp\left(\frac{t_0-t}{\tau_R}\right)
\end{align}
$t_0$ is the time of occurrence of an event which is the object of the
retention, $t$ is the present moment ($t > t_0$); $\tau_R$ is the characteristic
time associated to the decrease of the retention as we move away form the
occurrence of the event. Notice that when $\tau_R$ tends to $0$, $R(t_0,t)$ tends to $0$. $a_R$ is a coefficient which can be associated to an
individual or to a species, for example, in comparison to others of which such
faculties are more or less developed.

We propose to use relaxation functions, because the loss of retention, by moving
away from the moment of the beginning of a phrase or, more generally, from the
beginning of any action (including listening), can be considered as a sort of
“return to equilibrium”. A necessary return if we want to grasp the meaning of
the ensuing phrase or action. This, obviously, does not preclude us from
maintaining a memory of a more long-term past (the initial part of a discourse,
for instance): we limit ourselves to an analysis of the local, pre-conscious
effect which contributes to the extended present of an ongoing activity.

How may we now formally define virtual protention, a property which belongs only
to living phenomena? We propose to make it mathematically intelligible by means
of a \emph{temporal symmetry} with regard to $R$ (time $t$ will change sign). 
So we define, by a symmetry adjusted by two new parameters, $a_P$  and 
$\tau_P$, a \emph{virtual} protention. Now, time $t_1$ is the time of the event
to be anticipated and which is in the future of the present instant
$t$  ($t_1 > t$),  in the form of the function:
\begin{align}
V_{P}(t,t_1) &= a_P\exp\left(\frac{t-t_1}{\tau_P}\right)
\end{align}
Where the different parameters, $a_P$ and $\tau_P$, play the same \emph{mutatis
mutandis} role as those which intervene in $R$ (cf fig. A). In particular, $\tau_P=0$ leads to  $V_{P}(t,t_1)=0$.

Finally, we  define  \emph{protention} $P(t,t_0,t_1)$ by the product $R V_{P}$:
\begin{align}
P(t,t_0,t_1) &= R(t_0,t) V_{P}(t,t_1)  = a_P a_R\exp\left[\frac{t_0-t}{\tau_R}\right]\exp\left[\frac{t-t_1}{\tau_P}\right]\label{procap}
\end{align}
The (linear) dependance of $P$ on $R$, according to the principles stated above,
emphasizes that such a capacity can only
exist, phenomenologically speaking, if there exists in one form or another a
sort of “memory” $R$ (retention) relative to the event of which the reiteration
or something resembling it is to be anticipated (we are aware that we are making
a strong but empirically plausible hypothesis here, see footnote \ref{maning}).
In our view,  the specific traits of
this “expectation” of an unknown future, protention, is not exactly symmetrical
with
regard to the retention of a known past. And this by the fact that
protention
depends on retention -- and not conversely -- and that, by its nature, it
remains “potential” (it is the expectation of a ``possible'' event).

In the case where $R = 0$ (complete absence of retention), the protention is
 cancelled out by the fact that there no longer exists any referent
enabling to anticipate the expected event.

Still from the phenomenological standpoint, we will expect that in general $
\tau_P \ll \tau_R$, that is, that the characteristic time of retention be
greater than that associated to protention $P$ (in order to
“anticipate”, it is first necessary to “remember”, as stressed above). So the
contribution of
$V_{P}$ in the definition of $P$ (the second exponential in $\tau_P^{-1}$),
evolves more rapidly than that of retention for a same concerned duration. And
we will always have $P \leq a_P R$, as a function of time $t$, and this for
any values of $\tau_P$ and $\tau_R$  ($P = a_P R$  being achieved only in the
very moment that the time to be anticipated is the actual present, that is for 
$t = t_1$  and hence  $\exp\left[\frac{t-t_1}{\tau_P}\right] = 1$ ).

To make the role of the parameter $t$ more explicit, with regard to the interval
 $(t_0,t_1)$  and to the characteristic times  $\tau_P$, $\tau_R$,  some
simple algebraic manipulations enable to put the expression $P$ in the form of
the product of a function of $t$ and of two coefficients solely dependent on $t_0$
and $t_1$,  that is:
\begin{align}
P(t) &= a_Ra_P\exp\left[\frac{\tau_R - \tau_P}{
\tau_R\tau_P}(t-t_0)\right]\exp\left[\frac{t_0-t_1}{
\tau_R}\right]\exp\left[\frac{(\tau_R - \tau_P)}{
\tau_R\tau_P}(t_0-t_1)\right]\label{cap}
\end{align}

\begin{figure}[htbp]
\centering
\includegraphics[scale=0.8]{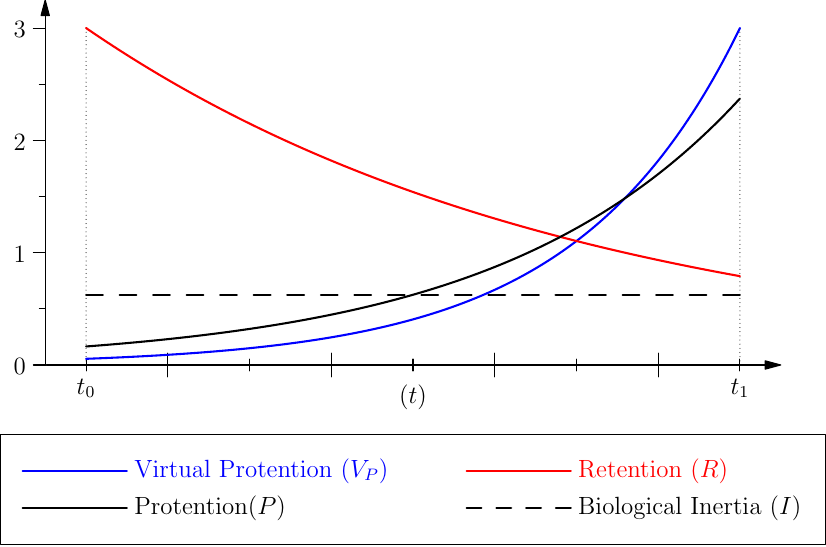}
\caption[Illustration]{\emph{Illustration of the basic quantities we define.}
Notice that protention is a growing function of time.}
\label{fig:graph3}
\end{figure}

\subsection{Comments}
\begin{figure}[htbp]
\centering
\includegraphics[scale=0.8]{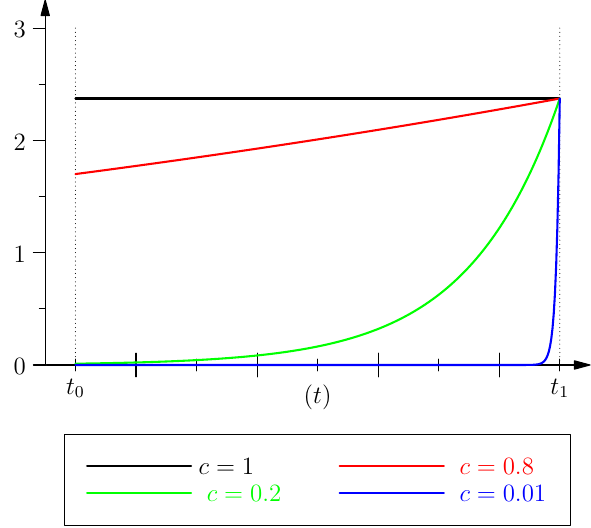}
\caption[Protention]{\emph{Protention} for various values of the ratio
$c=\frac{\tau_P}{\tau_R}$. We observe that small value of $c$ leads to a sharp
curve near $t_1$ whereas value close to $1$ are flat in the interval. We will
discuss the biological meaning of this case in section \ref{bioinert}.}
\label{fig:graphb}
\end{figure}

First, we should notice that $\frac{\tau_R\tau_P}{(\tau_R - \tau_P) }$ is an interesting quantity: it has the dimension of a time and is  the caracteristic time of $P(t)$.\\
 When $\tau_P$ tends to $\tau_R$, this quantity tends to infinity, and respectively  
$\frac{(\tau_R - \tau_P)}{\tau_R\tau_P}$ tends to $0$. This means that when $\tau_P$ is close to $\tau_R$,  $P(t)$ is almost stationary as a function of $t$. \\
On the contrary,when $\tau_R \gg \tau_P$, minor changes in time strongly
affect $P(t)$. More precisely, $P(t)$ is small when far from $t_1$ (and close to $t_0$), while it is very sensistive to (small) changes of $t$, when $t$ is close to $t_1$.  This means that, in this condition, the vicinity of the virtual event is where
the effect of protention is important, see figure \ref{fig:graphb}.\\ 
It is crucial, however, to understand that
protention, for example in the case of a cognitive situation, is not
empirically associated with a change of behaviour, but with the \emph{speed} of
this change of behaviour. This suggests a way to approach these quantities
empirically by a comparison of the reaction time between situations where
the event associated with retention (at time $t_0$)  occurs and when it \emph{doesn't}: in the first case, a more sudden change is then to be expected close to the the expectation time $t_1$.
Alternatively, the situation when the event at time $t_0$ occurs but where the event (at
time $t_1$) doesn't occur allow to evidenciate the presence of
protention and to see a part of its effects, it is the case of amoeba in \cite{amoeba}. However, in many situations, the effect of protentional action will consist in a
``sensibilization'' to the virtual stimuli with the preparation of a
response. This may lead to no behavioural change when the virtual stimuli
doesn't happen, but leads to a change of organization associated with the preparation of the response (including at the sensory level) and possibly to a greater sensitivity to noise.

\subsection{Global protention}
\noindent One may wonder when protention is
maximal for a given individual.
In our approach, the first possible answer is given by looking at the diagram in
figure \ref{fig:graphb}: this quantity is maximal close to $t_1$. However, we
can refine the
question (and the answer) by looking at the global amount of protention along the intended interval $[t_0, t_1]$. As protention is both variant and contravariant in the size of $[t_0, t_1]$ (see definition
\ref{procap}), this
question has a non-obvious answer.

For this purpose, we define the notion of
\emph{global protention}, which is the sum (the integral) of protention
 over time, between $t_0$ and $t_1$.   

\begin{align}
G_P(t_1 -t_0)&=\int_{t_0}^{t_1}P(t)dt \\
& = \frac{a_Ra_P \tau_R\tau_P}{\tau_R -
\tau_P} {\exp\left[\frac{t_0\tau_P - t_1\tau_R
}{\tau_R\tau_P}\right]}\left(\exp\left[\frac{(\tau_R - \tau_P)}{
\tau_R\tau_P}t_1\right]-\exp\left[\frac{(\tau_R - \tau_P)}{
\tau_R\tau_P}t_0\right]\right)\\
 & = \frac{a_Ra_P \tau_R\tau_P}{\tau_R - \tau_P} \left(\exp\left[\frac{t_0 - t_1
}{\tau_R}\right]-\exp\left[\frac{t_0 - t_1 }{\tau_P}\right]\right)
\end{align}
\begin{figure}[htbp]
\centering
\includegraphics[scale=0.6]{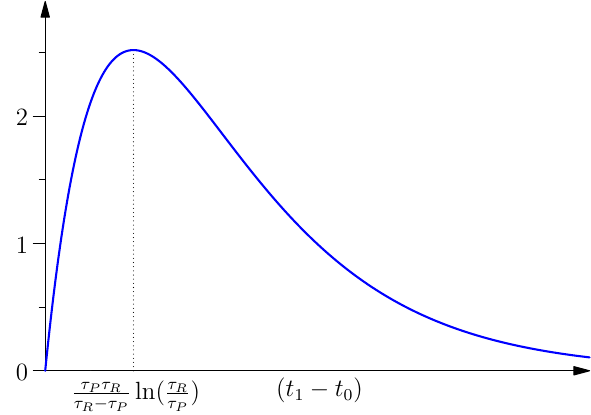}
\includegraphics[scale=0.6]{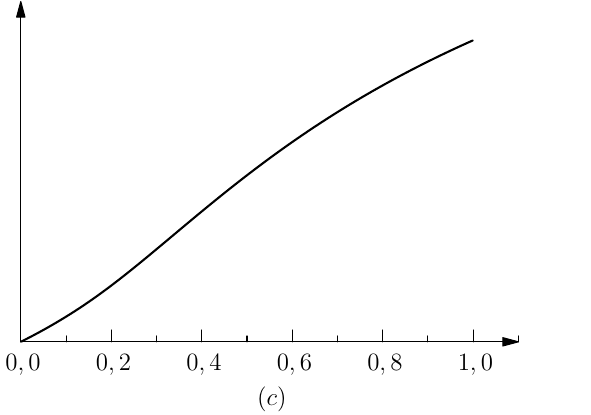}
\caption[Global protention]{\emph{Global protention}. When considered as a
function of the length of the  time interval (LEFT), there is a maximum
which corresponds to the greater effect of the couple Protention/Retention.
RIGHT, we see the global protention as a function of $c=\frac{\tau_P}{\tau_R}$.
}
\label{fig:graph2}
\end{figure}
This quantity has a maximum for $t_1-t_0=\frac{\tau_P \tau_R}{
\tau_R-\tau_P}\ln(\frac{\tau_R}{ \tau_P})$, this maximum is a compromise between
the need to give the protention time to have effect (covariant
dependence on the size of $[t_0,t_1]$) and the need to
have instants in $[t_0,t_1]$ that are close \emph{both to} $t_0$ and $t_1$
(contravariance). This result
means that there is a specific duration between the past event and the future
 event which optimize the protentional effects. This seems to be
consistent with the results in \cite{amoeba}, since these authors found that a
specific value of the delay $t_1 - t_0$ (in our notation) leads to a greater
protentional effect, that is the functional dependency on this interval of time
has a maximum (a non-obvious fact). In section \ref{bioinert} we will
go back to the relevant ratio $c=\frac{\tau_P}{\tau_R}$.

\section{Biological inertia}
\label{bioinert}
\noindent Consider now a relaxation phenomenon in physics, typically given
by $\Phi(t) = d\exp\left[\frac{t_0-t}{\tau_R}\right]$. If time $t_1 > t_0$ is
given, one may decompose $\Phi(t)$ as
\begin{align}
 \Phi(t) &= d\exp\left[\frac{t_0-t_1}{\tau_R}\right]
\exp\left[\frac{t_1-t}{\tau_R}\right]
\end{align}
The coefficient, not depending on $t$, that is
$d\exp\left[\frac{t_0-t_1}{\tau_R}\right]$, is the
``residual'' at time $t_1$ and it may be understood as a form of ``inertia'' of
the intended relaxed quantity (for example, it corresponds to ``what remains''
at time $t_1$ of a compound which decay with characteristic time $\tau_R$). This
coefficient is constant in the interval and decreases for increasing $t_1$.

In eq. (\ref{cap}) one has the following factors that do not depend on
$t$: 
\begin{align}
 a_Ra_P\exp\left[\frac{t_0-t_1}{\tau_R}\right]\exp\left[\frac{(\tau_R - \tau_P)}{\tau_R\tau_P}(t_0-t_1)\right]
\end{align}

The first exponential term corresponds  to a physical inertia, let's call
it $I_\varphi(t_0,t_1)$. Then, we can consider that the other coefficient
of protention represents a 
\emph{biological inertia}, in the interval $[t_0,t_1]$, depending on the
biological constants $a_R$, $a_P$, $\tau_R$ and $\tau_P$:
\begin{align}
I(t_0, t_1)=a_Ra_P\exp\left[\frac{(\tau_R -
\tau_P)}{\tau_R\tau_P}(t_0-t_1)\right]
\end{align}

In other words,
protention in eq. (\ref{cap}) may be considered as a product of a function of
time $t$, $\exp\left[\frac{\tau_R - \tau_P}{
\tau_R\tau_P}(t-t_0)\right]$, modulated by constants and characteristic times,
of a physical inertia $I_\varphi(t_0,t_1)$ and of a  “biological inertia”
$I(t_0,t_1)$. This last
coefficient is also independent of $t$, but depends on the specific organism
by the various indexed constants. 

The physical inertia represents the
``passive''
decay of a physical relaxation phenomena, which makes a perturabtion disappear
during the return to equilibrium. On the contrary, the biological inertia
coefficient is to be understood as a capacity to “carry over” the
protensive effect. Their names are freely inspired by the inertial mass as a
coefficient of acceleration (thus and very informally, biological inertia would
be the biologically pertinent coefficient of protention). In section
\ref{bio-inertia}, by references and a discussion, we will say more about this
new concept. First a few technicalities.\\

We have to check whether our definitions depend on the specific reference we
choose. That is to say if a time origin change: \\
\begin{align} t_0 & \leftarrow \tilde{t}_0=t_0 +\Delta t & t_1 & \leftarrow \tilde{t}_1=t_1 +\Delta t& t & \leftarrow \tilde{t}=t +\Delta t\end{align}
changes the way we split $P$ in three parts, in equation \ref{cap}. It
it then straightforward to see that:
\begin{align}
 \exp\left[\frac{\tau_R - \tau_P}{ \tau_R\tau_P}(t-t_0)\right] &=\exp\left[\frac{\tau_R - \tau_P}{ \tau_R\tau_P}(\tilde{t}-\tilde{t_0})\right] \\
 \exp\left[\frac{t_0-t_1}{ \tau_R}\right] &= \exp\left[\frac{\tilde{t_0}-\tilde{t_1}}{ \tau_R}\right]\\
 a_Ra_P\exp\left[\frac{(\tau_R - \tau_P)}{ \tau_R\tau_P}(t_0-t_1)\right]&=a_Ra_P\exp\left[\frac{(\tau_R - \tau_P)}{ \tau_R\tau_P} (\tilde{t_0}-\tilde{t_1})\right]
\end{align}
This means that each of this quantities have a sound biological meaning.

Inertia introduces a coefficient which is independent of $t$ and
is, in general, much smaller than $a_Ra_P$ (and always smaller than $a_Ra_P$). This
coefficient contributes to the dependence of $P$ in function of $t$. In
particular, it contributes in an essential manner to the decrease of the
protention according to the temporal distance.

\subsection{Analysis}

\begin{figure}[htbp]
\centering
\includegraphics[scale=0.65]{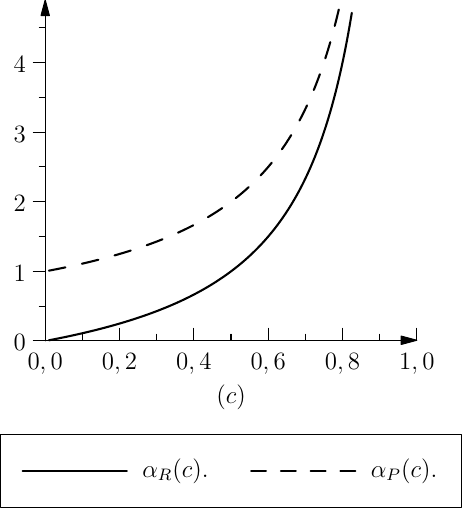}
\includegraphics[scale=0.65]{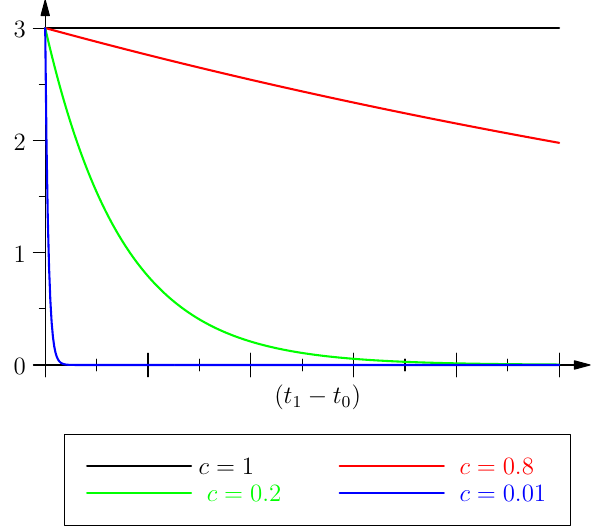}
\caption[Biological inertia]{ \emph{Biological inertia}. LEFT: we plot the factor of the characteristic time  of biological inertia seen as protention (or retention).  RIGHT: biological inertia as a function of the length of the time interval for various values of $c$.  }
\label{fig:graphd}
\end{figure}

In order to better understand the sense we attribute to this “inertia” of living
phenomena, given our preceding remark regarding orders of magnitude respective
of characteristic times, we may focus on the ratio $c$ of figure
\ref{fig:graph2}, that is on $c$ such that $\tau_P = c \tau_R $. We 
consider $0 \leq
c \leq 1$ and rewrite $I$ in the equivalent form:

\begin{align}
I( t_1-t_0) &=a_Ra_P \exp\left[\frac{(1 - c)}{c \tau_R}(t_0-t_1)\right] &\label{pr} \\
&=a_Ra_P \exp\left[\frac{1}{\alpha_R(c) \tau_R}(t_0-t_1)\right]& \text{ with }\alpha_R(c)=\frac{c}{(1 - c)}
\end{align}

Then $I$ has the \emph{form} of a “long term retention” if $c>0.5$ or a ``short term retention'' if $c<0.5$. 
Conversely, and maybe even more intuitively, inertia can be also interpreted (by
writing $\tau_R = \frac{\tau_p}{c}$ and eliminating this time $\tau_R$) as a
“long term virtual protention” :

\begin{align}I( t_1-t_0) &=a_Ra_P \exp\left[\frac{(1 - c)}{ \tau_P}(t_0-t_1)\right] & \label{pp}\\
&=a_Ra_P \exp\left[\frac{1}{\alpha_P(c) \tau_P}(t_0-t_1)\right]& \text{ with }\alpha_P(c)=\frac{1}{(1 - c)}
\end{align}

Biological inertia would then be both an extended retention, eq. (\ref{pr}), and
a virtual
protention, eq. (\ref{pp}), which are both \emph{independent} of the time $t$ of the
action: in fact,
it depends only on the instants that are relevant to the event retained and
occurring in $t_0$ or which is the object of an expectation (protention towards
$ t_1$).  It is therefore an inertia which “carries over” the life form from
$t_0$ towards $t_1$, by the preservation of its own
structure and its relationship with the environment (see section 
\ref{bio-inertia}).\\

\textbf{The $\tau_R = \tau_P $ case}\\

\noindent It can be observed that in the case where the characteristic retention
and
virtual protention times are equal ($\tau_R = \tau_P = \tau$ where the $c$ from the
equation above is equal to 1), the protention $P$ becomes
$a_Ra_P\exp\left[\frac{(t_0-t_1)}{\tau}\right]$ and is therefore independent
of the present observation time $t$. This, of course, within the interval between
the moment of the occurrence of the event in question and the moment $t_1$ where
it is mobilized again (since we still have $t_0 < t < t_1$ ).  But then,
according still to hypothesis $ c = 1$, one has  $P =
a_Ra_P\exp\left[\frac{(t_0-t_1)}{\tau}\right]$, with 
$I(t_0,t_1) =
a_Ra_P$.  Thus, when $ (\tau_R = \tau_P)$,
 only inertia is present in protention.\\

We can also note this situation by considering that if the observation time $t$
is close to the instant $t_0$ of the occurrence of the event (recent retention),
then the temporal interval for a virtual protention, $(t-t_1) \approx
(t_0-t_1$), increases; conversely, if time $t$ is far from $t_0$  (remote
retention), the temporal interval involved in this virtual protention and within
which the latter plays its role (the future of the observation moment $t$)
diminishes in importance, given of course that the protention $P$
as such remains independent of $t$, in this case.

These remarks are meant to
highlight the fact that, in the latter case, the \emph{intensity} $P$ of the
protention remains invariant, whereas the duration upon which
virtual protention takes place --- the future of $t$ --- can change in size: 
$t_1 - t$.

\section{References and more justifications for biological inertia}
\label{bio-inertia}
\noindent We have come to propose a mathematical notion of biological inertia
through an
apparently arbitrary play of symmetries and calculations, of which we would now
like to better explain the meaning and the objectives. To emphasize the
importance of the concept, but without wanting to make excessive and
uncontrolled analogies with immensely illustrious precedents, let’s note that
modern physics started off with a good analysis of inertia, as a ``pursuing a
state'' without aim nor teleology: Galilean
inertia\footnote{Without forgetting Giordano Bruno who had an informal yet quite
relevant notion of inertia, a few years prior to Galileo. It then
became possible to understand planetary movements without God being required to push
the planets around at all times. We similarly aim at a concept of inertia for
living phenomena with no reference to “vital impetus” or divine thrust.}.

In biology, this notion can already be found, although rarely, under various
forms. For \cite{vaz1978} “the lymphoid system has an \emph{inertia}, which
resists attempts to induce sudden and profound deviations in the course of
events”. So this is a weak notion of inertia, close to the “persistence” of
structural stability. Likewise, we could talk about inertia in the case of the
notion of “dynamic core” presented in \cite{edelman2000}, because it also refers
to the continuity/persistence of individuation (see also \cite{levanquyen2003}).
This theme is also used by \cite{varela1997}, where the term of inertia appears
also in the attempt to grasp the “force”, specific to any organism, enabling its
“bringing forth of an identity”.

In our approach, which is inspired by the methods of physics without identifying
with it, we firstly define retention by a relaxation function, which is a
physical notion --- which can even be considered as adequate to describe the
“memory” to which some often refer in relation to certain physico-chemical
activities. Virtual protention  is given then by a temporal symmetry,
\emph{modulo} some adjustment coefficients; this notion, which has no analogy in physics, is by this, and at least, the “projective” reflection of retention.
Protention follows, as a linear combination of these two values, in
function of time. Then, by a simple algebraic device, we separate the part
containing the temporal variable from the functional definition: what remains is
a constant, a function of all other parameters (characteristic times, specific
constants, interval range), which we called biological inertia. As we
pointed out, when retention and protention have the same characteristic
times $(\tau_R = \tau_P)$, inertia coincides with  protention. We
would then say that this is the simplest situations from a
cognitive viewpoint: the organism can only anticipate by means of inertia. In any
case, the proposed notion of inertia appears to clearly specify the informal
idea of “bringing forth of an identity”, with the reference to retention and to
protention, at the minimal cognitive level.

But why would this inertia not simply correspond to the fact of following a
geodesic trajectory, like in physics? Some will say that the amoeba, the
paramecium, etc., follow a gradient in the same way that a physical object
follows the trajectory dictated by the Hamiltonian (through the principle of
least action). It may appear that such is the case in \emph{in vitro}
experiments where, within a highly purified environment, the unicellular
organism is exposed to one or two very specific gradients (chemical,
thermal~\ldots). On the other hand, in an \emph{in vivo} situation, in the
ecosystems preferred by such animalcules (and which are very polluted, from our
standpoint) they must “arbitrate” between qualitatively different stimuli:
several physico-chemical gradients, an edible and close bacterium that is not
too large, another smaller one, etc. Now the paramecium, say, appears to “learn”
(see [Mislin, 2004]), that is, it enjoys at least retention, which contributes
to protention (and, after reading Mislin and references, one could posit for it 
$\tau_R > \tau_P$,  or even  $\tau_R \gg \tau_P)$\footnote{A paramecium manages
the movements of about 2,000 cilia during highly complex swimming activities;
some of its cilia also serve to direct food towards a “mouth” (opening upon the
membrane), by means of very articulate movements.}. And it is difficult to
conceive of learning without error, or without several attempts and without the
memory of these attempts (retention), even if such memory is extremely
rudimentary. The subsequent action is therefore one among many possible ones,
from the standpoint of the ecosystem, because it also depends on the specificity
of individual retention (experience). Among these many possible trajectories,
the one it follows has only to be \emph{compatible} with the ecosystem. No
gradient or physical geodesic is adequate to describe this plurality of
possibilities of evolution, phylogenesis, ontogenesis and of action, which also
depends on the specificity, hence on the history, of the species or of the
individual (retention and biological inertia). Our modest inertial attempt tries
to do this, in a way that is as preliminary as mathematically simple.

We can interpret the growth of  $
(\tau_R - \tau_P) \geq 0$ as a greater cognitive “complexity”.  It appears
that the protention, when
$\tau_R \gg \tau_P $,  must account for more “experience” in order to achieve
the objective of the action; it depends upon a greater amount of lived and
retained history, and hence on a greater specificity (individuality) of the
living object. So it better participates to the incessant process of
individuation, which is a play between the richness of retention and the
diversity of possible future trajectories.\\

Another way to associate a growth of complexity to the growth of  $
(\tau_R - \tau_P) \geq 0$, is to consider cases where the \emph{global}
protention is constant. Then the increase of $
(\tau_R - \tau_P) \geq 0$ means that protention is more localized near $t_1$,
with the same global effect. Then this situation is more ``complex'', since
the preparation to the virtual event occurs when it is closer (and the organism
must be ``quickly ready''). In this case, it is easier for it to
protend another event $t_1'$, with $t_1'$ between $t_0$ and $t_1$, since the
organism is not yet fully focused on $t_1$ (the $P$ grows very slowly
``for long'' and fastly increases only close to $t_1$). This situation allows
the organism to have longer times of correlation: during the early part of these
extended protentional activities, it may prepare also for other events .

\section{Towards human cognition. From trajectory to space: The continuity of
the
cognitive phenomena}

\noindent The continuity of space-time, which the mathematics of continua
proposes and
structures in a remarkable way, from Euclid to Cantor,
follows --- and does not precede --- the continuity of a figure, of a contour or
of a trajectory. Euclidean geometry is not a geometry of space, it is a geometry
of figures, with continuous edges, constructed by means of ruler and compass
and submitted to translations and to rotations. It is much later, with
Descartes, that geometry finds its
constitutive environment in an abstract space, underlying and independent from
the figures which evolve within. The analytical reconstruction of Euclidean
geometry will follow, by means of this ideal framework, an
algebraico-geometrical continuum, organized in Cartesian coordinates. Then, since Cantor, we have a fantastic reconstruction of the underlying continuum, a
possible one, though (see Lawvere and Bell for an alternative
topos-theoretic approach, with no points, \cite{Bell98} ).

Let’s now try to grasp a possible constitutive path or even a cognitive
foundation of this \emph{phenomenal} continuum which is the privileged
conceptual and mathematical
tool for the intelligibility of space, on the basis of our analysis of retention
and of protention.

The recent analyses of the primary cortex (see \cite{Petitot08} for a
survey)
highlight the role of intracortical synaptic linkages in the perceptual
construction of edges and of trajectories. Neurons correlate themselves
locally, along ``association fields'' (\cite{Field87, Field93}) composed of
smooth (differentiable) curves that ``are grouped toghether only when alignement
fails along particular axes'' \cite{Field93}. These neurons are sensitive
to “directions”: that is, they
activate when detecting a direction, along a tangent. Then they (pre-)activate
other neurons in the association field (they prepare in advance the spike
which is not yet fired). This preactivation of associated neurons is, in our
view, a component of the protensive activity. Then, neuronal activation follows
a specific direction which (re-)constructs the pertinent line, \cite{Petitot08}.

Thus, the continuity
of an edge or of a trajectory is constructed by “gluing” together fragments of
the world, in the
precise geometrical
(differential) sense of gluing.  In other words, we force \emph{by continuity}
the unity of an edge by relating neurons which are pre-associated and
are, locally, along particular axes.

This phenomenon participates in the retention and the protention of a
\emph{non-existing} line, a trajectory say, by ``integrating'', in the mathematical
sense, the tangents that are locally associated in the field. The related
inertial
 phenomena of the activation/deactivation of neurons may be one of its
constitutive elements, with inertia as a coefficient of protention. The
retention of occular movements or saccades which follow a moving body, an edge,
should also be quoted: this retentive/protensive phenomenon originates
in the muscles enabling the saccades or in the neurons managing them. As for the
case of protention, in particular, there are
protentional displacements in the receptor field of the cortical neurons that
\emph{precede} the saccades (\cite{Berthoz02}). The brain
prepares itself and
anticipates a moving object, of which the movement is perceived following an
occular saccade, or of which the trajectory or edge is perceived by running the
eye along or over it.
This is, in our view, the keystone of a fundamental protentional activity.

Now, we propose the following conjecture. First, the World is not
continuous, nor discrete: it is what it is. Since Newton and Cantor, by
continuous tools, or, now, in Quantum
theories and Topos Theoretic approaches, we mathematically organized it in
various ways, possibly over different ``backgrounds''. In our view, the
phenomenal continuity of trajectories, of an edge, is due to
the \emph{retention} of that trajectory, edge, scanned by the eye, which
is \emph{“glued”} with the \emph{protention} by the very unit of the cerebral and global physiological activity (the vestibular system, for example, has its own
retention and inertia). 

In the case of contours, the specific saccades along
the direction of movement or towards the extreme of a reconstructed segment 
(for example in Kanizsa triangles, see \cite{Petitot08}) stimulates a specific
activation in the association field (a specific connection between neurons in
the field). 

It would then be this “gluing” --- a mathematically
solid concept (at the center of differential geometry, of which Riemannian geometry is a special case) --- that entails the
cognitive effect which \emph{imposes} continuity upon the world: the image of
the object and of its past position is reassembled (glued by the conjunction of
protention and retention) with that of the object and of its expected position
or a contour is made continuous even when non existing (as in Kanizsa
illusions).
We could indeed imagine that an animal with no fovea (the part of the eye which
enables a follow up of a target by a continuous focus), a frog for example, and
which takes spaced out snapshots of an object in movement would not
have the impression of a continuous movement in the way in which we,
the primates, “see” it.

By measuring relaxation and (pre-)activation times of associated neurons
it should be possible to quantify our coefficients in these specific phenomena. 
Inertial coefficients in particular would yield different values according to
the different protentional capacities in different species (frogs for example
may have no inertia w.r. to these phenomena, if our understanding above is
correct).

So the continuity of a trajectory or of an edge is, in our
opinion, the result of a spatio-temporal reassembling of the retentions and
protentions that are managed by global neural activity in the presence of a
plurality of activities of such type (muscles, vestibular system~\ldots but also
the differentiable continuity of the movement or gesture participates by means
of its own play of retention/protention). In short, by a cognitive
process of glueing, we attribute continuity to phenomena which are what they
are (and which a frog surely sees quite differently). Then, by a remarkable
conceptual and mathematical effort having required centuries, we have even come to theorize, as abstract lines, surfaces and their edges, first, and then even the continuity of environing space, as the background of these structures. And this is the consequence, we believe, not the cause of the
cognitive/perceptive continuity of the movement and of the gesture, which is instead grounded on the unity of protention and retention (note that, in this perspective, the continuity of an edge would also be the
continuity of a movement: the movement of the saccade or of the hand caressing
it, both retained and protended).

Let's note that, in our attempt towards spatialization of time for living
phenomena, in this paper and in \cite{bailly2010},
--- a spatialization which, although schematic, should contribute to its
intelligibility --- we have proceeded, in this section, along the opposite
approach: a sort of temporalization of space. Its apparent continuity would be
the result of a cognitive activity \emph{on} time, the extended present
obtained by rentention and protention.

\paragraph{Acknowledgements}
We would like to thank the anonymous referee for his/her critical comments
and suggestions. Longo's papers are downloadable from
\url{http://www.di.ens.fr/users/longo/}. 

\bibliographystyle{spbasic}
\bibliography{bib}

\end{document}